\documentclass[11pt, twopage, reprint, superscriptaddress,  preprintnumbers,nofootinbib, amsmath, amssymb]{article}

\usepackage{mathrsfs}
\usepackage{amsfonts}
\usepackage{graphicx}
\usepackage{blindtext} 
\usepackage{bookman}
\usepackage[T1]{fontenc}
\usepackage{microtype}
\usepackage[english]{babel}
\usepackage[hmarginratio=1:1,top=32mm,columnsep=20pt]{geometry}
\usepackage[hang, small,labelfont=bf,up,textfont=rm,up]{caption}
\usepackage{booktabs}
\usepackage{lettrine}
\usepackage{enumitem}
\setlist[itemize]{noitemsep}
\usepackage{abstract}
\usepackage{titlesec}
\titleformat{\section}[block]{\large\scshape\centering}{\thesection.}{1em}{}
\titleformat{\subsection}[block]{\large}{\thesubsection.}{1em}{}

\usepackage{fancyhdr}
\usepackage{titling} 
\usepackage{hyperref}
\hypersetup{
    colorlinks=true,
    linkcolor=blue,
    filecolor=magenta,      
    urlcolor=blue,
    citecolor=blue,
    pdftitle={Electromagnetic instantons and asymmetric Hawking radiation of black holes},
    pdfpagemode=FullScreen,
    }

\usepackage[dvipsnames]{xcolor}
\usepackage[affil-it]{authblk}
\usepackage{amssymb,physics,slashed}
\usepackage{graphicx}
\usepackage{upgreek}
\usepackage{cite}

\title{Electromagnetic instantons and \\ asymmetric Hawking radiation of black holes}
\author{
\textsc{Archil Kobakhidze and Elden Loomes} 
\vspace{0.2cm} \\
\normalsize \itshape
Sydney Consortium for Particle Physics and Cosmology, \\
\normalsize  \itshape
School of Physics, The University of Sydney, NSW 2006, Australia \\
\normalsize \itshape \href{mailto:archil.kobakhidze@sydney.edu.au}{archil.kobakhidze@sydney.edu.au}, 
\normalsize \itshape \href{mailto:elden.loomes@sydney.edu.au}{elden.loomes@sydney.edu.au}, 
}
\date{}
\renewcommand{\maketitlehookd}{%

\begin{abstract}
\noindent 
We argue that the topological structure of Abelian gauge theories, such as Maxwell electrodynamics, in the background of a Euclidean Schwarzschild black hole manifests itself through an asymmetry in Hawking radiation. In particular, the topology of the black hole manifold, characterised by a non-contractible 2-sphere and Euler characteristic $\chi = 2$, admits non-trivial gauge-field configurations. These take the form of 2-form field strengths that are closed but not exact. From a topological perspective, such configurations are classified by the second cohomology group, which is isomorphic to $\mathbb{Z} \oplus \mathbb{Z}$, and are labelled by integer electric ($n$) and magnetic ($m$) charges, $(n,m)$. Self-dual ($n = m$) and anti-self-dual ($n = -m$) dyonic configurations carry vanishing Euclidean energy and are fully compatible with the Euclidean Schwarzschild geometry. More general dyonic configurations, by contrast, are interpreted as off-shell Euclidean field configurations. Nevertheless, both classes contribute to the thermal equilibrium vacuum and to finite-temperature correlation functions in the corresponding Lorentzian framework. Furthermore, because of the non-trivial topology, the electromagnetic $\uptheta_{\rm EM}$-term contributes to the physical observables. In particular, it sources $CP$-asymmetric Hawking radiation, observable as an imbalance between left- and right-polarised photons in the emission spectrum. We briefly discuss some implications of this phenomenon.  
         
\end{abstract}
}

\begin{document}
\maketitle

\section{Introduction}

According to the black hole uniqueness theorems, black holes formed via gravitational collapse settle into equilibrium states uniquely defined by their mass, angular momentum, and charge \cite{Ruffini:1971bza}. During black hole evaporation \cite{Hawking:1975vcx}, only these conserved quantities remain accessible to distant observers; all other information about the initial matter is lost. For instance, a non-rotating black hole (with zero angular momentum) is expected to emit chirally symmetric Hawking radiation, that is, photons of both helicities in equal amounts, yielding a net zero spin-angular momentum in the radiation. 
      
In this paper, we argue that even the simplest non-rotating, neutral black hole possesses a rich topological structure, arising from the non-trivial gauge-field configurations supported by its geometry. In particular, the thermal vacuum surrounding a Schwarzschild black hole includes integer-valued electric and magnetic dyonic states, associated with harmonic 2-form gauge field strengths that are closed but not exact. Within this framework, the topological $\uptheta_{\rm EM}$-term acts as a source of chirally asymmetric Hawking radiation.  

While the phenomenon we describe applies universally to any 0-form gauge theory, we focus on Maxwell electrodynamics in the background of a static, non-rotating, neutral black hole in order to highlight the subtle interplay between spacetime geometry and topology. In flat spacetime, the vacuum of Maxwell theory is topologically trivial. However, when coupled to gravity, the electrodynamic vacuum can acquire non-trivial topological structure, potentially giving rise to observable effects \cite{Deser:1980kc,Arunasalam:2018eaz}. In particular, we show that, in a black hole background, Maxwell gauge fields develop topological features that profoundly modify the thermal vacuum and lead to a charge-parity ($CP$) asymmetry in Hawking radiation, manifested as an imbalance in the emission of left- and right-polarised photons.

Our discussion is structured as follows. In the next section, we review Maxwell field configurations in the background of Euclidean black holes and explore their topological properties. In Sec.~\ref{three}, we discuss the physical interpretation of the topological dyonic solutions. In Sec.~\ref{four}, we argue that the topological electromagnetic $\uptheta_{\rm EM}$ becomes a physically observable parameter, leading to $CP$-violating effects, most notably, asymmetric Hawking radiation of photons. Finally, Sec.~\ref{five} is devoted to a summary and discussion.     

\section{Maxwellian topological charges of a Euclidean Schwarzschild black hole}
\label{two}
\subsection{Euclidean Schwarzschild black hole}
The thermal features of black holes are described within the Euclidean formalism with the Wick-rotated imaginary time coordinate, $\tau$. The metric for the Euclidean Schwarzschild manifold, $\mathcal{M}  \cong \mathbb{R}^2\times S^2$, reads:
\begin{align}
    \mathrm{d} s^2 = \Big(1-\frac{2GM}{r}\Big)\mathrm{d} \tau^2 + \Big(1-\frac{2GM}{r}\Big)^{\!-1}\mathrm{d} r^2 + r^2(\mathrm{d}\theta^2+\sin^2\theta\,\mathrm{d} \phi^2)~.
\end{align}
These coordinates cover a subset $(\mathbb{R}^2\backslash\{r=2GM\})\times S^2 \subset \mathcal{M}$ of the Schwarzschild manifold. If the correlation functions evaluated on such a background manifold are to be interpreted as the analytic continuation of the correlation functions on the physical Lorentzian spacetime, the horizon singularity $r=r_{\rm h}\equiv 2GM$, must be eliminated.  This is done by compactifying the imaginary time coordinate $\tau$, with period $\beta =4\pi r_{\rm h}$, which is exactly the reciprocal of the Hawking temperature, $T_{\rm H}$. Correlators computed in the background of the Euclidean Schwarzschild manifold describe equilibrium thermal correlators of the Lorentzian black hole, in the limit where the backreaction of the thermal radiation on the black hole geometry can be neglected. This corresponds to the double scale limit: $G\to 0$, $M\to \infty$, with $r_{\rm h}$ fixed. 

The topological features of manifolds are characterised by two topological invariants, the Euler characteristic, $\chi$, 
\begin{equation}
\chi \equiv\sum_n(-1)^nb_n = \int_{\mathcal{M}}{\rm Tr}\,R \wedge \tilde R+\text{(bound. terms)}~,
\label{euler}
\end{equation}
and the Hirzebruch signature, $\tau$,
\begin{equation}
\tau \equiv b_2^+-b_2^{-}=\int_{\mathcal{M}}{\rm Tr}\, R\wedge R + \text{(bound. terms)}~.
\label{signature}
\end{equation}
Here $\tilde R_{\mu\nu\rho\sigma}=\frac{1}{2}\epsilon_{\mu\nu\alpha\beta}R^{\alpha\beta}~_{\rho\sigma}$, is the dual Riemann tensor and $b_n$ is the $n$th Betti number (the number of non-contractable $n$-surfaces). The Euclidean Schwarzschild manifold contains a non-contractible 2-sphere, $S^2$, located at the tip $r=r_{\rm h}$ of the manifold. Thus, the Euler characteristic is $\chi=2$, and the manifold admits a harmonic 2-form that can wrap around $S^2$. This will be discussed next.

\subsection{Topology of $L^2$ harmonic 2-forms}
Consider Maxwell's electrodynamics on the Euclidean Schwarzschild manifold, which is described by a closed $\mathfrak{u}(1)$ $2$-form field strength, $F$, $\mathrm{d} F=0$. Its classical motion is governed by the Maxwell action,
\begin{align}
    S_\mathrm{EM}&=\frac{1}{2e^2}\int_\mathcal{M} F\wedge *F~, \label{eq:Maxwell action}
\end{align}
for which the equations of motion are $\mathrm{d} *F=0$. Thus $F$ is a harmonic form. The classical solutions for $F$ are thus classified by Hodge theory, which for a manifold $\mathcal{M}$ gives a correspondence between harmonic forms $\mathcal{H}^\bullet(\mathcal{M})$ on $\mathcal{M}$ and the cohomology groups $H^\bullet(\mathcal{M})$.

Recall that every exact $k$-form $\omega=\mathrm{d} \gamma$, is closed, $\mathrm{d} \omega=0$, but not every closed $k$-form is globally exact. Rather, each closed form may only be exact on a a number of topologically trivial local neighbourhoods $\{U^i\}$ covering $\mathcal{M}=\bigcup_i U^i$, $\omega\big|_{U^i}=\mathrm{d} \gamma^i\big|_{U^i}$, where each $\gamma^i$ differs by closed forms on their overlap. The failure of this exactness is measured by the $k$th cohomology group,
\begin{align}
    H^k(\mathcal{M})\equiv \frac{\ker\mathrm{d}|_\text{$k$-forms}}{\mathrm{im}\,\mathrm{d}|_\text{$(k-1)$-forms}}~.
\end{align}
In the case of the field strength $F$, its closure means that it may be written locally as the derivative of a $\mathfrak{u}(1)$-valued 1-form potential $A$, $F=\mathrm{d} A$. For field strengths not in the trivial class of $H^2(\mathcal{M};\mathbb{Z})$, the closed form is not exact and hence cannot be represented as $\mathrm{d} A$ globally. That is, more than one patch is required to cover $\mathcal{M}$. On the boundary between patches, the 1-form potentials must then differ by a local gauge transformation (transition function), $A^i-A^j=g^{-1}\mathrm{d} g=\mathrm{d} \alpha$, for $g=e^{i\alpha}\in U(1)$ and $\alpha$ defined modulo $2\pi$. The prototypical example is the Wu--Yang construction of a monopole \cite{Wu:1975es, Wu:1976ge}, where the classes of $H^2(\mathcal{M};\mathbb{Z})$ are labelled by the magnetic charge.

An important modification to the topology is introduced by requiring that the action \eqref{eq:Maxwell action} is finite. This is critical for being able to perform semiclassical calculations. As the action is proportional to the $L^2$ norm $\lVert\omega\rVert_{2}=\int \omega\wedge *\omega$, we must focus on $L^2$($-$normalisable)  harmonic 2-forms $L^2\mathcal{H}^2(\mathcal{M};\mathbb{Z})$. When $\mathcal{M}$ is asymptotically locally flat, then the asymptotic decay of $F$ means that it is sufficient to study $\mathcal{M}$ compactified with a single point at infinity, $\hat{\mathcal{M}}\equiv \mathcal{M}\cup \{\infty\}$ \cite{Hausel:2002xg}.

\begin{figure}[t]
\centering
\includegraphics[width=0.5\textwidth]{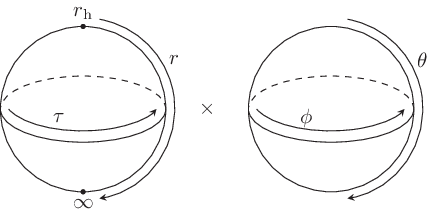}
\caption{The topology of the compactified Euclidean Schwarzschild manifold, $S_{(\tau,r)}^2\times S_{(\theta,\phi)}^2$.}
    \label{fig:compactification}
\end{figure}

When compactified with a point at infinity, the Euclidean Schwarzschild manifold becomes $\hat{\mathcal{M}}\cong  S^2_{(\tau,r)}\times S^2_{(\theta,\phi)}$, where we label the spheres with their coordinates for clarity. See Figure \ref{fig:compactification}. The space of $L^2$ harmonic 2-forms on the compactified Euclidean Schwarzschild manifold is then  isomorphic to the second cohomology group of this compactification,
\begin{align}   {L^2}\mathcal{H}^2(\mathcal{M};\mathbb{Z})\cong H^2(S^2_{(\tau,r)}\times S^2_{(\theta,\phi)};\mathbb{Z})\cong \mathbb{Z}\oplus \mathbb{Z}~,
\end{align}
labelled by the integer charges $(n,m)$, counting the windings around the two 2-spheres.
Explicitly, the $L^2$ harmonic 2-forms on $\mathcal{M}$ are
\begin{align}
    F=\frac{n}{2r^2}\mathrm{d}\tau\wedge \mathrm{d} r +\frac{m}{2}\sin\theta\,\mathrm{d}\theta\wedge \mathrm{d} \phi ~.
\end{align}
These are dyons, with electric and magnetic charges $n$ and $m$, respectively.\footnote{These solutions have been first discussed in \cite{Charap:1977re} within $SU(2)$ gauge theory. Later, it has been shown in \cite{Etesi:2000mv} that those solutions display, in fact, entirely Abelian characteristics.} These charges are measured by integration over any closed Gaussian 2-surface $\Sigma$ wrapping around $S_{(\theta,\phi)}^2$:\footnote{For example, $\Sigma=\{(\tau_0,r_0)\}\times S_{(\theta,\phi)}^2$ for some fixed $(\tau_0,r_0)\in S_{(\tau,r)}^2$}
\begin{align}
    n=\frac{1}{2\pi}\int_\Sigma *F\,,\qquad m=\frac{1}{2\pi}\int_\Sigma F~.
\end{align}
Since,
\begin{align}
    *\Big(\frac{1}{r^2}\mathrm{d}\tau\wedge \mathrm{d} r\Big)=\sin\theta\,\mathrm{d}\theta\wedge \mathrm{d} \phi~,
\end{align}
we can alternatively define,
\begin{align}
    n=\frac{1}{2\pi}\int_T F\,,\qquad m=\frac{1}{2\pi}\int_T *F~,
\end{align}
where the Gaussian surface $T$ now wraps around $S_{(\tau,r)}^2$.\footnote{For example, $T= S_{(\tau,r)}^2\times \{(\theta_0,\phi_0)\}$ for some fixed $(\theta_0,\phi_0)\in S_{(\theta,\phi)}^2$} Clearly, electric and magnetic charges swap under the electromagnetic duality transformation. 

As discussed, the closed harmonic forms $F$ are not exact, that is, they cannot be represented as $\mathrm{d} A$ everywhere, since the gauge potentials $A$ are singular. Following \cite{Wu:1975es, Wu:1976ge}, one defines patches on the manifold that support non-singular gauge potentials $A$. At the boundary of such patches, $A$ is multi-valued, the difference given by a local gauge transformation. The single-valuedness of the boundary gauge transformation then implies topological quantisation of the respective charge, described as windings of field configurations. 

The winding around $S^2_{(\theta,\phi)}$ proceeds exactly as in the case of the Wu--Yang construction \cite{Wu:1975es, Wu:1976ge}. We cover $S_{(\theta,\phi)}^2$ with a pair of patches $H^\pm$ consisting of the Northern and Southern hemispheres respectively, and denote $\hat{\mathcal{M}}^\pm = S_{(\tau,r)}^2\times H^\pm$. The patches overlap at the equator, where $\hat{\mathcal{M}}^+\cap \hat{\mathcal{M}}^-\cong S^2_{(\tau,r)}\times S^1_{(\phi)}$. The difference between 1-form potentials $A^\pm$ supported on each patch must be a local gauge transformation $A^+-A^-=\mathrm{d}\alpha$, whose winding determines the integer magnetic charge, $m\in\mathbb{Z}$. This is the standard Dirac quantisation. 

On the compactified Euclidean Schwarzschild manifold, we can perform a similar construction to argue for the quantisation of an electric charge. We cover $S_{(\tau,r)}^2$ with a pair of patches: an infinitesimal closed neighbourhood $X^\infty$ of the point at infinity ($r\to \infty$) and $X^{r<\infty}=\overline{S_{(\tau,r)}^2\backslash X^\infty}$, the rest of the sphere.\footnote{Alternatively, we could choose a different patching with the boundary at the horizon, $r=r_{\rm h}$. These two patchings result in the same physics.} Then denote $\hat{\mathcal{M}}^{r<\infty}= X^{r<\infty}\times S^2_{(\theta,\phi)}$ and $\hat{\mathcal{M}}^{\infty}=X^\infty \times S^2_{(\theta,\phi)}$ supporting gauge fields $A^{r<\infty}$ and $A^{\infty}$ respectively, and overlapping on $\hat{\mathcal{M}}^{r<\infty}\cap \hat{\mathcal{M}}^\infty\cong S_{(\tau)}^1\times S_{(\theta,\phi)}^2$. The potential on the patch at infinity must then match with the bulk potential by a local gauge transformation $A^{\infty}-A^{r<\infty}=\mathrm{d} \alpha '$. Similar to the case of magnetic charges, the single-valuedness of such gauge transformations leads to the dual Dirac quantisation of electric charges, $n\in \mathbb{Z}$.    
For configurations with both electric and magnetic charges, we must use both sets of patches at once, breaking the compactified spacetime into four parts: $\hat{\mathcal{M}}^{{r<\infty},\pm}=X^{r<\infty}\times H^\pm$ and $\hat{\mathcal{M}}^{\infty,\pm}=X^\infty\times H^\pm$ with a gauge transformation at each boundary.

Explicitly, a valid choice of 1-form potentials on the Euclidean Schwarzschild manifold is: 
\begin{align}
 A^{r<\infty}_{\tau}&= -\frac{n}{2r_\mathrm{h}}\Big(1-\frac{r_\mathrm{h}}{r}\Big)\,,\quad A^{\infty}_{\tau}=0~, \\
A_{\phi}^\pm &=\frac{m}{2}(\pm 1-\cos\theta)~,
\end{align}
where $+(-)$ corresponds to the gauge potential on the Northern (Southern) hemisphere of $S_{(\theta,\phi)}^2$ and $A^{r<\infty}_{\tau}$ is supported everywhere except of at radial infinity. 

At the equator $\theta=\pi/2$, $A_{\phi}^{\pm}$ differ by the local gauge transformation: 
\begin{align}
    A_{\phi}^+-A_{\phi}^-=\mathrm{d}\alpha\,,\qquad \alpha=m\phi~.
\end{align}
Requiring $g=e^{i\alpha(\phi)}=e^{i\alpha(\phi+2\pi)}$ to be single-valued forces $m$ to be integer, $m\in \mathbb{Z}$. This is precisely the Wu--Yang construction of integer magnetic charges.    

At infinity, $r\to \infty$, the gauge transformation is
\begin{align}
    A^{\infty}_{\tau}-A^{r<\infty}_{\tau}=\mathrm{d}\alpha'\,,\qquad \alpha'=\frac{n}{2r_\mathrm{h}} \tau=\frac{2\pi n}{\beta}\tau~.
\end{align}
Once again, requiring the gauge transformation to be single-valued forces $n$ to be an integer, $n\in \mathbb{Z}$. This is the dual Dirac quantisation of an electric charge. 

Closing this section, we point out that since the compactified Euclidean Schwarzschild manifold supporting $L^2$ harmonic forms is $S^2\times S^2$, $b_2^{+}=b_2^{-}=1$ and thus $b_2= b_2^{+}+b_2^{-}=2$. In the absence of other non-trivial harmonic forms \cite{Etesi:2000mv}, all other Betti numbers vanish, while compactness ensures no boundary terms contribute in Eqs.~(\ref{euler}) and (\ref{signature}). Hence, $\chi=2$ and $\tau=0$.    

\section{Physical interpretation}
\label{three}
It is well established that the Euclidean black holes contributing as saddle-point classical solutions to the effective Einstein-Hilbert action result in a thermal partition function describing the system in thermal equilibrium at the Hawking temperature \cite{Gibbons:1976ue}. Here, we would like to elaborate on the physical interpretation of dyonic configurations introduced in the previous section. 

For a particular $(n,m)$ dyonic configuration, the Maxwell action reads:  
\begin{align} S_\mathrm{EM}&=\frac{2\pi^2}{e^2}(n^2+m^2)~,
\label{action}
\end{align}
while its Euclidean energy-momentum tensor takes the form:
\begin{align}
    T_{\mu\nu}=(m^2-n^2)t_{\mu\nu}~,
\label{em}
\end{align}
where the non-zero tensor components are:
\begin{align}
 {t}_{\tau\tau}=\Big(1-\frac{r_\mathrm{h}}{r}\Big)^2\,,\qquad t_{rr}= 
 \frac{1}{2 r^4}\Big(1-\frac{r_\mathrm{h}}{r}\Big) \,,\qquad {t}_{\phi \phi }=\sin^2\theta\,,\qquad  t_{\theta\theta}=-\frac{\sin ^2\theta }{2 r^2} ~.
\end{align}
Note, that only self-dual ($n=m$) and anti-self-dual ($n=-m$) dyonic configurations with vanishing Euclidean energy-momentum tensor (\ref{em}) are compatible with the classical Euclidean Schwarzschild solution. Thus, they contribute as saddles to the partition function.

More general dyonic solutions at the classical level are compatible with magnetically and electrically charged Euclidean black holes. However, quantum mechanically, they contribute as off-shell configurations to the Euclidean path integral in the background of a neutral Schwarzschild black hole. Factoring out the pure gravity part, the partition function in the saddle-point approximation,  
\begin{align}
Z\simeq\sum_{n,m}\,e^{-S_\mathrm{EM}}~,
\label{pf}
\end{align}
describes a statistical ensemble of dyons carrying $(n,m)$ topological charges. The ensemble average of an observable $\mathcal{O}$ is then computed as:
\begin{align}
\langle\mathcal{O}\rangle&\simeq\frac{1}{Z}\sum_{n,m}\mathcal{O}\,e^{-S_\mathrm{EM}}~.
\label{eq:saddle approximation}
\end{align}
It should be stressed that only these expectation values/correlators can be faithfully Wick rotated back to describe features of the Lorentzian black hole.

At this point, it is useful to comment on a technical aspect. Within the semiclassical approximation, the contribution of the dyon ensemble to the partition function (18) and the corresponding correlators (19) is determined entirely by a discrete sum over classical solutions. This stands in contrast to the standard instanton calculus, which requires integration over moduli spaces that characterise the positions and sizes of instantons. In the present case, however, the solutions are associated with a black hole of fixed position and size, and as a result, translational and scale invariance are explicitly broken.

The above picture, in fact, is fully compatible with the semiclassical description of gravity. Indeed, the expectation value of the energy-momentum tensor and the black hole charge vanish, 
\begin{align}
\langle T_{\mu\nu}\rangle &\simeq \frac{1}{Z}\sum_{n,m} T_{\mu\nu}\,e^{-S_\mathrm{EM}}= \frac{t_{\mu\nu}}{Z}\sum_{n,m} (m^2-n^2)\exp\!\Big[-\frac{2\pi^2}{e^2}(n^2+m^2)\Big] =0\,,\label{eq:energy-momentum}\\
 \langle n\rangle &\simeq \frac{1}{Z} \sum_{n,m}n\,e^{-S_\mathrm{EM}} =\frac{1}{Z}\sum_{n,m} n\exp\!\Big[-\frac{2\pi^2}{e^2}(n^2+m^2)\Big]=0~.\label{eq:electric_charge}
\end{align} 
Similarly, $\langle m\rangle =0$ for an ensemble-averaged magnetic charge. This implies that the off-shell contribution from non-(anti-)self-dual solutions is compatible with a classical neutral Schwarzschild black hole within the semiclassical treatment. Furthermore, the semiclassical approximation is always valid in the thermal equilibrium regime, defined by the double scaling limit mentioned earlier. 

An analogy between flat spacetime instantons at finite temperature and $(n,m)$ dyons becomes more transparent by inspecting the Pontryagin charge. 
Because $F$ splits into orthogonal electric and magnetic parts $F=F_e(r,\tau)+F_m(\theta,\phi)$, supported on the two 2-spheres, $F\wedge F=2F_e\wedge F_m$, and the Pontryagin charge $P$ factorises into the product of quantised electric and magnetic charges,
\begin{align}
    P&\equiv\frac{1}{8\pi^2}\int_\mathcal{M} F\wedge F=\frac{1}{4\pi^2}\int_\mathcal{M} F_e\wedge F_m=\Big(\frac{1}{2\pi}\int_{S_{(\tau,r)}^2} F_e\Big)\Big(\frac{1}{2\pi}\int_{S_{(\theta,\phi)}^2} F_m\Big)=nm~.\label{eq:Pontryagin number}
\end{align}
We see that this indeed represents an integer topological charge. 
The action (\ref{action}) for a $(n,m)$ dyon then satisfies 
the Bogomolny--Prasad--Sommerfeld bound,
\begin{align}
    S_\mathrm{EM} \geq \frac{4\pi^2}{e^2}|P|~,
\end{align}
This bound is saturated precisely by the (anti-)self-dual $n=|m|$ dyons.

It is instructive to push this analogy further. At finite temperature, instantons become periodic in compact Euclidean time and admit multi-centre solutions, known as calorons \cite{Harrington:1978ve}. When the instanton size $\rho$ exceeds the thermal scale $\beta$ ($\rho \gg \beta$), thermal over-the-barrier transitions dominate, while quantum-mechanical tunnelling is exponentially suppressed. In this regime, the caloron solution transitions into a solution represented by static dyons that collectively carry the topological charge. Therefore, the change in topological charge is interpreted as a fluctuation in the statistical ensemble of dyons rather than quantum tunnelling events. 

Our dyonic configurations admit a natural interpretation along similar lines to calorons in the high-temperature regime. The effective ``size'' of the dyons is set by $r_{\rm h}$. Being localised near the horizon, their scale should be compared with the local temperature,
$T_{\rm loc} = T_{\rm H}\left(1 - \frac{r_{\rm h}}{r}\right)^{-1}$. Since $r_{\rm h} \gg 1/T_{\rm loc}$ as $r \to r_{\rm h}$, the system is effectively always in the high-temperature regime, where topological fluctuations are dominated by thermal effects. 

We are thus led to the following physical picture. The Euclidean Schwarzschild black hole carries topological charges $(n,m)$ associated with dyonic configurations. These charges are intrinsically quantum and cannot manifest directly in a classical setting. They rather come in the form of a statistical ensemble of dyons which contribute to the thermal equilibrium state of a semiclassical Schwarzschild black hole. The topology change is therefore best understood as arising from fluctuations in the ensemble of dyons.  In what follows, we discuss a potentially observable imprint of this modified equilibrium state on Hawking radiation.
\section{Electromagnetic $\uptheta_{\rm EM}$ and asymmetric Hawking radiation}
\label{four}
The addition of the topological $\uptheta_{\rm EM}$-term,
\begin{equation}
S_{\uptheta_{\rm EM}}= -\frac{i\uptheta_{\rm EM}}{8\pi^2}\int_\mathcal{M} F\wedge F
\label{theta}
\end{equation}
to the Maxwell action (\ref{eq:Maxwell action}) in flat spacetime has no physical consequences, as the $U(1)$ gauge group is topologically trivial. This situation changes once magnetic monopoles are introduced alongside electric charges. In that case, the 
second cohomology group becomes non-trivial, forcing the gauge group to be compact. This leads to the Dirac quantisation condition, whereby electric charges are quantised in units set by the monopole charge. Moreover, the $\uptheta_{\rm EM}$-term turns monopoles into dyons, endowing them with electric charge proportional to $\uptheta_{\rm EM}$ \cite{Witten:1979ey}. As a result, $\uptheta_{\rm EM}$ becomes a physically observable parameter.

In curved spacetime, the situation is further enriched: depending on the topology of the underlying manifold, $\uptheta_{\rm EM}$ can become physical even in pure Maxwell theory, without the need to introduce external charges \cite{Deser:1980kc, Arunasalam:2018eaz}. In particular, since dyons in the background of the Euclidean Schwarzschild black hole carry non-zero Pontryagin charge (\ref{eq:Pontryagin number}), one expects $\uptheta_{\rm EM}$ to be observable in this setting.

Indeed, evaluating the ensemble-averaged Pontryagin charge and retaining only the leading contributions from configurations with $(n,m)=(\pm1,\pm1)$ and $(\pm1,\mp1)$, we find a non-vanishing result:
\begin{align}
    \langle P\rangle 
    &\simeq \frac{1}{Z}\sum_{n,m} nm \, e^{-S_\mathrm{EM}} \simeq 2e^{-4\pi^2/e^2+i\uptheta_{\rm EM}} - 2e^{-4\pi^2/e^2-i\uptheta_{\rm EM}} \nonumber \\
    &\simeq 4i\, e^{-4\pi^2/e^2}\sin\uptheta_{\rm EM}\,.
    \label{top}
\end{align}
Note that the factor $i$ is absorbed upon Wick rotation back to Lorentzian signature, so the physical topological charge is real. This non-zero expectation value of the topological charge serves as a topological order parameter for $CP$ symmetry breaking. At the same time, the ensemble averages of the energy-momentum tensor (\ref{eq:energy-momentum}) and the electric and magnetic charges (\ref{eq:electric_charge}) remain zero, so the semiclassical description of the Euclidean Schwarzschild black hole is unaffected.  

We will now argue that the non-zero topological charge (\ref{top}) is directly responsible for a non-zero photon helicity flux, implying an imbalance between left- and right-handed circularly polarised photons in Hawking radiation. The 4-current density that accounts for such flux is given by (at least locally) \cite{Dolgov:1988qx}: 
\begin{equation}
    J_{\rm h}=*\left(A\wedge F\right)\,,
    \label{cur}
\end{equation}
This is known as the magnetic helicity current density in the context of plasma physics. Its divergence,  
\begin{equation}
    {\rm d}*J_{\rm h}=F\wedge F\,,
    \label{div}
\end{equation}
is given (up to a numerical factor) by the Pontryagin charge density (\ref{eq:Pontryagin number}).\footnote{In addition to the classical non-conservation, $J_{\rm h}$ exhibits quantum anomaly in the gravitational background with non-zero $R\wedge R$ \cite{Dolgov:1988qx}. Since the Hirzebruch signature (\ref{signature}) is zero in our case, the gravitational quantum anomaly does not apply.}

For a given $(n,m)$ dyon we find that helicity density $J_{\rm h}^0=0$, while the radial current density $J_{\rm h}^r=
\frac{1}{\sqrt{g}} A_\tau F_{\theta\phi}=
-\frac{1}{r^2}\frac{nm}{2r_{\rm h}}\Big(1-\frac{r_{\rm h}}{r}\Big)$ depends on the corresponding Pontryagin charge $P=nm$. This results in a non-zero ensemble-averaged photon helicity flux at infinity
\begin{align}
\Phi_{\rm h}=\int_{r\rightarrow\infty}\! \langle J_{\rm h}^r\rangle\,r^2{\rm d}\Omega_2 = 
-{32\pi^2 }i\,T_\mathrm{H}e^{-4\pi^2/e^2}\sin\uptheta_{\rm EM}\,.
\end{align}
The flux is real in Lorentzian spacetime. This, \emph{in thesi}, observable feature in Hawking radiation arises from the non-trivial topological structure of Maxwell theory in the background of a non-rotating, neutral black hole. 

\section{Summary, discussion and outlook}
\label{five}

In this paper, we have argued that the Maxwell gauge field possesses a non-trivial topological structure in the background of a Euclidean Schwarzschild black hole, which gives rise to $CP$ violation and induces an asymmetry in the Hawking radiation of static, non-rotating, neutral black holes. This topological structure is encoded in the harmonic 2-forms on the compactified Euclidean Schwarzschild manifold. Compactification follows from the requirement that the configurations have finite action, changing the topology from $\mathbb{R}^2\times S^2$ to $S^2\times S^2$. These forms are classified by their winding numbers around the 2-spheres $S_{(\tau,r)}^2$ and $S_{(\theta,\phi)}^2$, which are interchanged under the electromagnetic duality. The corresponding integer windings $(n,m)$ represent quantised electric ($n$) and magnetic ($m$) charges of dyons, forming a statistical ensemble that contributes to the thermal equilibrium state of a semiclassical Schwarzschild black hole. In this framework, topological fluctuations are realised as fluctuations within the dyon ensemble.

These dyonic configurations support a non-trivial electromagnetic $\uptheta_{\rm EM}$-term, thereby rendering the $\uptheta_{\rm EM}$ parameter physically observable. In particular, we have shown that although the ensemble-averaged electric and magnetic charges vanish, the average topological Pontryagin charge remains non-zero. This result is directly related to a non-vanishing, $CP$-violating helicity flux of thermal photons in the Hawking radiation.

We further note that, within the semiclassical approximation, the contribution of the dyon ensemble to the partition function and the associated correlators is given entirely by a discrete sum over classical solutions. In contrast to the standard instanton calculus, which involves integration over moduli spaces characterising the position and size of instantons, our solutions are tied to a black hole of fixed position and size. Consequently, translational and scale invariance are explicitly broken. 

We envisage several interesting directions for further investigation of the phenomenon discussed in this paper. These are briefly outlined below.

\subsection{On the origin of $\uptheta_{\rm EM}$}

A realistic theory is considerably richer than the pure Maxwell theory considered here for illustrative purposes. While we have introduced $\uptheta_{\rm EM}$ by hand, it is instructive to examine how such a term may arise naturally within more complete theoretical frameworks. For instance, if electrodynamics is non-trivially embedded in a theory based on a simple gauge group, as in grand unified theories, then at low energies it inherits an effective $\uptheta_{\rm EM}$ from the underlying fundamental theory. In such settings, the electromagnetic $\uptheta_{\rm EM}$ parameter is related to the more familiar $\uptheta_{\rm QCD}$ of quantum chromodynamics. Moreover, since the vacuum structure is globally constrained by the full grand unified gauge symmetry, one generically expects additional topological features to emerge in the low-energy effective theory.

An effective $\uptheta_{\rm EM}$ may also arise in a black hole background through the pion--photon coupling~\cite{Kobakhidze:2008em}, induced by the Adler--Bell--Jackiw anomaly~\cite{Adler:1969gk,Bell:1969ts}. Furthermore, it can be generated radiatively in the presence of chiral matter, for example, in cosmological scenarios involving a chiral primordial plasma (see, e.g.,~\cite{Barrie:2017mmr}).

\subsection{Inclusion of fermions}
The inclusion of fermions coupled to gauge fields is known to dramatically modify the non-perturbative vacuum structure. In an instanton background, fermions exhibit zero modes of the Dirac operator that carry net anomalous charge, in accordance with the topological index theorem \cite{index,Atiyah:1975jf}. As recently discussed in  \cite{Dvali:2024dlb, Dvali:2024zpc, Dvali:2025pcx}, such vacuum zero modes can generically materialise in the form of composite fermionic condensates, which spontaneously break the anomalous symmetry and give rise to an emergent pseudo-Goldstone state. We expect a similar phenomenon to occur in a black hole background. This will be investigated elsewhere.

\subsection{Virtual black holes and cosmological black holes}
While the effect of an ensemble of dyons is vanishingly small for individual black holes, at least in weakly coupled regimes, it may become significant when the dynamics is dominated by a statistically large number of small black holes. There are two scenarios where we envisage this to happen.

It has been argued \cite{Hawking:1995ag} that the gravitational vacuum at Planckian scales is dominated by the production and subsequent evaporation of virtual black holes. In this picture, the vacuum is expected to carry additional topological features related to the dyonic solutions discussed in this work.\footnote{For comparison, Abelian instantons in charged wormhole backgrounds were constructed by \cite{GarciaGarcia:2025xqj}.} However, the effect at such short distances is expected to be more pronounced, as the quasiclassical exponential suppression is substantially reduced relative to its value in the large-scale regime. In particular, we expect the short-distance contribution from virtual black holes to the axion mass to be significant. If so, the effect may have implications for the cosmological evolution of axion dark matter \cite{Dvali:2026ceb}.  

Additionally, the early universe may have undergone a phase of primordial black hole domination \cite{Hooper:2019gtx}. Such small cosmological black holes would evaporate rapidly.\footnote{The evaporation rate of cosmological black holes may be modified \cite{Boehm:2020jwd} relative to that of isolated black holes due to the global expansion of the universe.} The asymmetric Hawking radiation during the evaporation phase may leave behind a net anomalous charge and/or polarised radiation, which could lead to observable signatures. 

In summary, there is a wide range of physical applications of the phenomenon discussed in this paper, which we plan to explore in future publications.

\section*{Acknowledgements}
Some aspects of this work were developed while the authors were visiting the Max Planck Institute for Physics and the Arnold Sommerfeld Centre for Theoretical Physics at Ludwig Maximilian University of Munich. We thank Gia Dvali and the members of his group, especially Lasha Berezhiani and Oto Sakhelashvili, for useful discussions and generous hospitality. EL would like to thank Ryuichiro Kitano for his valuable insight regarding fermion and axion effects, and the members of his group for their kind welcome throughout his visit to the Yukawa Institute for Theoretical Physics at Kyoto University. This work was partially funded by the Australian Research Council through the Discovery Projects grant DP220101721. EL also acknowledges support from the University of Sydney Grant-in-Aid and PRSS travel grants.

\end{document}